\documentclass[twocolumn,preprintnumbers,amsmath,amssymb,superscriptaddress]{revtex4-1}

\usepackage{graphicx}
\usepackage{dcolumn}
\usepackage{bm}
\usepackage{color}
\usepackage{float}

\begin{document}

\title{Disorder engineering: From structural coloration to acoustic filters}
\author{Nitin Upadhyaya}
\affiliation{School of Engineering and Applied Sciences, Harvard University, Cambridge, Massachusetts 02138, USA}
\affiliation{Center for mathematical modeling, Flame University, Pune, Maharashtra 412115, India}

\author{Ariel Amir}
\affiliation{School of Engineering and Applied Sciences, Harvard University, Cambridge, Massachusetts 02138, USA}

\begin{abstract}
We study Anderson localization of waves in a one dimensional disordered meta-material of bilayers comprising of thin fixed length scatterers placed randomly along a homogenous medium. As an interplay between order and disorder, we identify a new regime of strong disorder where the localization length becomes independent of the amount of disorder but depends on the frequency of the wave excitation and on the properties of the fixed length scatterer. As an example of a naturally occurring nearly one dimensional disordered bilayer, we calculate the wavelength dependent reflection spectrum for Koi fish using the experimentally measured parameters, and find that the main mechanisms for the emergence of their silver structural coloration can be explained through the phenomenon of Anderson localization of light in the regime of strong disorder discussed above. Finally, we show that by tuning the thickness of the fixed length scatterer, the above design principles could be used to engineer disordered meta-materials which selectively allows harmonics of a fundamental frequency to be transmitted in an effect which is similar to the insertion of a half wave cavity in a quarter wavelength stack. However, in contrast to the Lorentzian resonant peak of a half-wave cavity, we find that our disordered layer has a Gaussian lineshape whose width becomes narrower as the number of disordered layers is increased.  
\end{abstract}

\maketitle

Meta-materials are composite materials engineered out of more commonly available materials by carefully arranging them in ways such that their collective response gives rise to novel mechanical and electromagnetic properties \cite{book1, book2, book3, Pendry_2003,Cloak_2006,Alu_2004,Katia_2009,Fang_2011,Cook}. At the heart of any engineering design is the ability to accurately control the response of a system. Disorder thus seems manifestly at odds with the main principles of any engineering design. At the same time,  random aggregates of objects, both natural \cite{McKenzie_1995,Addadi_2010} or manmade \cite{Martin_2009,Nesterenko_2001,Ploog_2006}, often display many novel properties which emerge in part from their intrinsic disorder. Yet, characterizing disorder and harnessing it to design materials whose response can be precisely controlled remains a challenge.

A periodic arrangement of a bilayer of materials consisting of regions of two different wave speeds (mechanical or electromagnetic), but with randomly varying thicknesses, is a quintessential example of a one-dimensional disordered system whose transport properties are governed by Anderson wave localization \cite{Anderson_1958,Wiersma_1997,Baluni,Berry_1996,Makarov_2009, Anderson_book1, Anderson_book2}. Due to the mismatch in wave speeds at each interface of such a bilayer system, a part of the wave is reflected and a part gets transmitted. Disorder in the path lengths in a sufficiently large sample can then eventually cause the reflected waves to interfere constructively in such a way that all the energy remains confined to a region of space, known as the localization length. The localization length in general depends upon both the nature and amount of disorder and on the frequency of the incident wave. If the system length is much greater than the maximum localization length within the band of frequencies being considered, then wave transmission through the bilayer channel is effectively prevented and in the absence of any dissipative processes, all of the incident wave energy is reflected.

An intriguing prospect then is to ask the question: could the disorder-induced localization length be harnessed to {\it design} meta-materials whose frequency response can be precisely controlled? In this article, we study Anderson localization of waves in a one dimensional disordered meta-material consisting of thin fixed length scatterers placed randomly along a homogenous medium. We identify a different regime of wave localization whereby the localization length is independent of the amount of disorder (for sufficiently large disorder) yet depends only on the frequency of the wave excitation and on the properties of the fixed length scatterer. We use the transfer matrix formulation to analytically derive the localization length and power transmission and compare these to corresponding results obtained numerically. As an example of a naturally occurring nearly one dimensional disordered bilayer, we calculate the wavelength dependent reflection spectrum for Koi fish using experimental parameters discussed in Ref. \ (\cite{Addadi_2010}), and find that the main mechanisms for the emergence of their silver structural coloration can be explained through the phenomenon of Anderson localization of light in the regime of strong disorder discussed above.

Finally, we use our results to propose the design of a simple one dimensional mechanical meta-material and discuss how by tuning the wave speed and thickness of the scatterer, we can engineer disordered meta-materials which only allow the transmission of a narrow band of frequencies centered around the harmonics of the fundamental mode. We find that in contrast to resonances in ordered systems, which typically have Lorentzian lineshapes, here the transmission peak has a Gaussian shape, with a width that decays as $1/\sqrt{N}$, where $N$ is the number of layers, and thus can be made arbitrarily narrow.

\section{Anderson localization in a periodic on average disordered meta-material}

\begin{center}
\begin{figure}[h]
\includegraphics[width=.4\textwidth]{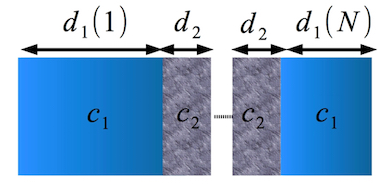}
\caption{\label{Schematic} A schematic illustration of a periodic on average disordered meta-material consisting of disordered bilayers of two materials with wave speeds $c_1,c_2$ respectively. The thickness of material 1, denoted by $d_1$ varies randomly while the thickness of material 2 is kept constant at a value $d_2$ which is much smaller than the average value of $d_1$. This is reminiscent of the random Kroning-Penny model whereby we place scatterers (with thickness $d_2$, wave speed $c_2$) randomly along the length of a homogenous background medium with wave speed $c_1$. For large variations in the thicknesses of material 1 (strong disorder),  the localization length for a wave with frequency $\omega$ depends only on the properties of the scatterer as derived in Eq.\ (\ref{Localization_Length_Zeta}). Thus, the parameters of material 2 can be tuned to design materials with a specific frequency response.}
\end{figure}
\end{center}

Consider a homogenous medium with scatterers of fixed length placed randomly along a medium, as in Fig.\ (\ref{Schematic}). This is reminiscent of the random Kroning-Penney model \cite{Makarov_PRB,Soukoulis_Book} where there is disorder in the thicknesses of only one of the layers (the background medium) comprising the bilayer while the thickness of the other is held constant at a value which is much smaller than the average thickness of the disordered layer.

In the following, we make use of the transfer matrix formulation to study the reflectance and transmittance of a wave (electromagnetic or acoustic) impinging on this one dimensional structure at normal incidence, though the approach presented generalizes to arbitrary angles of incidence. The transfer matrix which relates the forward going ($E^{+}$) and backward going $E^{-}$ complex wave amplitudes across a bilayer can be written in the form (see Appendix A for the derivation)
 \begin{flalign*}
\begin{pmatrix}
E^{+}_{\text{after bilayer}}\\
E^{-}_{\text{after bilayer}}\\
\end{pmatrix}=\mathcal{M}
\begin{pmatrix}
E^{+}_{\text{before bilayer}}\\
E^{-}_{\text{before bilayer}}\\
\end{pmatrix} \label{transMatrix}
\end{flalign*}
where,
\begin{flalign}
\mathcal{M}=\begin{pmatrix}
A & B \\
B^* & A^{*} \\
\end{pmatrix}
\end{flalign}
and
\begin{flalign}
A &=& \frac{1}{1-r^2}\left(e^{i(\delta _1+\delta _2)}-r^2e^{i(\delta _1-\delta _2)}\right),\\
B &=& \frac{2ir}{1-r^2}e^{-i\delta _1}\text{sin}\delta _2
\end{flalign}
where, $^*$ denotes complex conjugation and we have defined $r=\left|\frac{n_2-n_1}{n_2+n_1}\right|$ as the reflection coefficient for normal incidence at the interface between medium 1 and 2 with optical refractive indices $n_1,n_2$ respectively. Here, $\delta_{1,2}=\frac{2\pi}{\lambda}n_{1,2}d_{1,2}$ are the phases accumulated by a wave of wavelength $\lambda$ as it propagates medium 1 and 2 with thicknesses $d_{1,2}$ respectively, see Fig.\ (\ref{Schematic}).

We may rewrite the transfer matrix Eq.\ (\ref{transMatrix}) in the form $\mathcal{M}=\mathcal{M}_{s}\mathcal{M}_{b}$, where $\mathcal{M}_{s}$ is a completely deterministic transfer matrix associated with the fixed length scatterer, and $\mathcal{M}_{b}$ is a random transfer matrix associated with the background medium:
\begin{flalign*}
\mathcal{M}=\frac{1}{1-r^2}\begin{pmatrix}
e^{i\delta _2}-r^2e^{-i\delta _2} & -2ir\text{sin}\delta _2 \\
2ir\text{sin}\delta _2 &e^{-i\delta _2}-r^2e^{i\delta _2}\\
\end{pmatrix}
\begin{pmatrix}
e^{i\delta _1} & 0 \\
0 & e^{-i\delta _1}\\
\end{pmatrix}.
\end{flalign*}
Here, the entries in the second matrix ($\mathcal{M}_b$) contain only $\delta_1$ which is a random variable and denotes the phase accumulated by the wave as it traverses the background medium, while the entries in the first matrix ($\mathcal{M}_s$) contain only $\delta_2$ which is a constant phase change as the wave traverses the scatterer.

The transfer matrix across $N+1$ barriers then is recursively given by $\bold{M}_{N+1}=\mathcal{M}_{s}\mathcal{M}_{b}\bold{M}_N$, and assumes the form
\begin{flalign}
\bold{M}_{N+1}=\begin{pmatrix}
\frac{1}{\tau^* _1} & -\frac{\rho^* _1}{\tau^*_1} \\
-\frac{\rho _1}{\tau_1} &  \frac{1}{\tau _1}\\
\end{pmatrix}\begin{pmatrix}
\frac{1}{\tau^* _N} & -\frac{\rho^* _N}{\tau^*_N} \\
-\frac{\rho _N}{\tau_N} &  \frac{1}{\tau _N}\\
\end{pmatrix}
\end{flalign}
where, $\rho,\tau$ are complex numbers representing the wave amplitude reflected and transmitted respectively, see Eqs.\ (\ref{reflectance}) and (\ref{transmittance}). Thus, the power transmitted $T_{N+1} = \frac{1}{|\bold{M}_{N+1}[1,1]|^2}$ is
\begin{flalign}
T_{N+1} = \frac{T_NT_1}{1+R_NR_1+2\sqrt{R_nR_1}\text{cos}(\phi)} \label{T_N}
\end{flalign}
where, we have expressed the reflectance and transmittance amplitudes in terms of the power reflected $R$ and transmitted $T$ using the relations $\rho_1=\sqrt{R_1}e^{i\phi _1}$, $\rho _N=\sqrt{R_N}e^{i\phi _N}$, $\tau _N=\sqrt{T_N}e^{i\lambda _N}$, and defined $\phi=\phi _N-\phi _1-2\lambda _N$ to encode the random phase accumulated over the $N-$ layers.

Energy conversation and time reversal symmetry provide us with an additional constraint \cite{AJP}:
\begin{flalign}
\frac{\rho _N}{\rho^*_N}=-\frac{\tau _N}{\tau^*_N},
\end{flalign}
using which, $\lambda _N = \phi _N+\frac{\pi}{2}$ and therefore, $\phi$ assumes an even simpler form--
\begin{flalign}
\phi = \pi + \phi _1  + \phi _N. \label{Phi}
\end{flalign}
Eq. \ (\ref{Phi}) has a clearer physical interpretation-- $\phi$ is a random variable obtained by summing the reflection phases of individual bilayers through which the wave passes. The reflection phase of a single bilayer is obtained as $\rho = \frac{B^*}{A^*}$, see Eq.\ (\ref{reflectance}) and when $\delta _2$ is kept fixed, the reflection phase is $2\delta _1$ upto a constant. For strong disorder, i.e., when the standard deviation in the values of spacing $d$ between scatterers is very large compared with the wavelength, we would expect the phase $\delta _1 = 2\pi\frac{d}{\lambda}$ to become uniformly distributed within the interval $[-\pi,\pi)$. Adding a uniformly distributed phase to another phase characterized by any distribution would result in a uniformly distributed phase. Therefore in the strong disorder regime we expect $\phi$ to be uniformly distributed. As we will see in the following sections, even for moderate values of disorder in the strong disorder limit which we use in our work, the resultant distribution of $\phi$ tends to be uniform, a result which we will be able to further validate when we compare our analytical results with numerical data.

{Next, we take the logarithm of Eq.\ (\ref{T_N}) (which we shortly justify) and then, ensemble average both sides of the resultant equation \cite{Anderson_1958} to obtain:}
\begin{flalign}
\langle\text{log}T_{N+1}\rangle= \langle\text{log}T_{N}\rangle+\langle\text{log}T_{1}\rangle - \langle\text{log}(a+b\text{cos}(\phi))\rangle, \label{Log_T}
\end{flalign}
{where, $a=1+R_NR_1$ and $b=2\sqrt{R_nR_1}$. If we now use the condition that the probability distribution of $\phi$ is nearly uniform, the ensemble average of the last term in Eq.\ (\ref{Log_T}) vanishes as follows}:
\begin{flalign*}
\langle\text{log}(a+b\text{cos}(\phi))\rangle &=&  \frac{1}{2\pi}\int^{2\pi}_0\text{log}(a+b\text{cos}\phi)d\phi,\\
 								 &=&  \text{log}\left(\frac{1}{2}\left(a+\sqrt{a^2-b^2}\right)\right),  \\
								 &=&   \text{log}\left(\frac{1}{2}\left(1+R_NR_1+|1-R_NR_1|\right)\right),\\
								 &=&   0\label{Averaging}
\end{flalign*}
where, the last line follows since $|R_NR_1|\leq 1$.

{We now see one of the main advantages of taking the logarithm (see note \cite{Log_note_1} for further discussion) of Eq.\ (\ref{T_N}), for we have reduced Eq.\ (\ref{T_N}) into a simple recursion relation}:
\begin{flalign*}
\langle\text{log}T_{N+1}\rangle= \langle\text{log}T_{N}\rangle + \langle\text{log}T_{1}\rangle.
\end{flalign*}
whose solution is
\begin{flalign*}
\langle\text{log}T_N\rangle = N\langle\text{log}T_1\rangle.
\end{flalign*}
Substituting $T_1$ evaluated from Eq.\ (\ref{transMatrix}), we find that $T_1$ does not depend on the random variable $\delta _1$ but only on the fixed phase constant $\delta _2$ and hence is a deterministic quantity. The average of the log-power transmitted through $N$ bilayers is therefore:
\begin{flalign}
\langle\text{log}T_N\rangle                                            &=& -N\text{log}\left|\frac{1}{1-r^2}\left(e^{i\delta _2}-r^2e^{-i\delta _2}\right)\right|^2, \\
                                            &=& -N\text{log}\left(1+\frac{4r^2}{(1-r^2)^2}\text{sin}^2\delta _2\right). \label{Log_Avg_2}
\end{flalign}

{We can now invert the relation Eq.\ (\ref{Log_Avg_2}) to obtain the expected power transmitted through $N$ bilayers (see also \cite{Log_note_2}): }
\begin{flalign}
T_N \sim e^{-Nd/\zeta} \label{Power}
\end{flalign}
where $d=\langle d_1\rangle + d_2$ is the mean thickness of a bilayer and we obtain $\zeta$ as the localization length
\begin{flalign}
\zeta = \frac{d}{\text{log}\left(1+\frac{4r^2}{(1-r^2)^2}\text{sin}^2\delta _2\right)}.\label{Localization_Length_Zeta}
\end{flalign}
The resultant Eq.\ (\ref{Power}) can be expressed as $T_N \sim \frac{1}{\left(1+\frac{4r^2}{(1-r^2)^2}\text{sin}^2\delta _2\right)^N}$ or in other words, this suggests the power transmitted through the disordered layer can be obtained by taking the product of power transmitted through $N$ ordered bilayers, see Eq.\ (\ref{Power_Ordered}) for comparison. This is a remarkably simple result, and even though it may seem that the resultant transmittance is a result of incoherent transmission of waves, that is not the case. Using the results of Ref. \cite{Berry_1996} for \emph{incoherent} waves, one finds a Lorentzian lineshape whose width scales as $1/\sqrt{N}$. In turn, Eq.\ (\ref{Power}) gives a transmission peak which is \emph{Gaussian}, and whose width also happens to scale as $1/\sqrt{N}$. To appreciate this, consider Eqs. (\ref{Power}) and (\ref{Localization_Length_Zeta}). Defining $\Delta \equiv \frac{2r}{(1-r^2)}\text{sin}\delta_2$, the resulting transmission for a system with $N \gg 1$ layers and in the regions where $\Delta^2\ll 1$ is well approximated by:
\begin{flalign}
 T_N \approx e^{-N \text{log}(1+\Delta^2)} = \frac{1}{(1+\Delta^2)^N} \approx e^{-N \Delta^2}. \label{gaussian}
\end{flalign}

Further, we see from Eq.\ (\ref{Localization_Length_Zeta}) that under our assumptions of strong disorder, the localization length $\zeta$ is independent of the amount of disorder. However, since $\delta _2=\frac{\omega}{c_2}d_2$, the localization length depends on the frequency of the incident wave $\omega$ and on the parameters of the scatterer. In the next section, we first discuss the optical case and see how our result Eq.\ (\ref{Localization_Length_Zeta}) allows us to explain the observed reflection spectrum for Koi fish and consequently the main mechanisms for their silver structural coloration. This is one example of how nature harnesses disorder to engineer broadband reflectance. In the subsequent section, we will see how the localization length in Eq.\ (\ref{Localization_Length_Zeta}) provides us with a way to harness disorder and engineer the frequency response of an acoustic disordered meta-material.

What if we also allow the thickness of the scatterer to also vary randomly? In that case, $\delta _2$ in Eq.\ (\ref{Log_Avg_2}) is no longer a deterministic variable. However, if we continue to make the assumption that $\delta _2$ is uniformly distributed over $[0,2\pi)$ and express $2\sin^2(\delta _2)=1-\cos(2\delta _2)$, we can ensemble average Eq.\ (\ref{Log_Avg_2}) in the same way as  Eq.\ (\ref{Averaging}) to obtain
\begin{flalign}
\langle\text{log}T_1\rangle   = \frac{1}{2\pi}\int^{2\pi} _0\text{log}\left(1+\frac{2r^2}{(1-r^2)^2} -\frac{2r^2\text{cos}(2\delta _2)}{1-r^2}\right)d\delta _2, \nonumber \\
					  =  \text{log}\left(\frac{1}{1-r^2}\right). \label{Berry_Result}
\end{flalign}
This is exactly the result obtained in Ref. \cite{Berry_1996} where the authors explain the mirror like appearance of a stack of transparencies as a result of Anderson localization of light propagating through a one dimensional disordered medium composed of bilayers of air and plastic, both of whose thicknesses vary randomly.  Note here, the localization length in Eq.\ (\ref{Berry_Result}) depends only on the bare reflection coefficient $r$ and neither on the magnitude of disorder nor on the frequency of the incident wave. Consequently, such a meta-material will reflect back all incident frequencies and this is what gives a stack of transparent materials its characteristic mirror like appearance. In Appendix B, we re-derive our result Eq.\ (\ref{Localization_Length_Zeta}) using the method suggested in Ref.\ (\cite{Berry_1996,Baluni}) for the case when the thickness of one of the layers comprising the bilayer is held constant.

\section{Silver structural coloration in fish}

\begin{center}
\begin{figure}[h]
\begin{tabular}{c}
\includegraphics[width=4.4 cm]{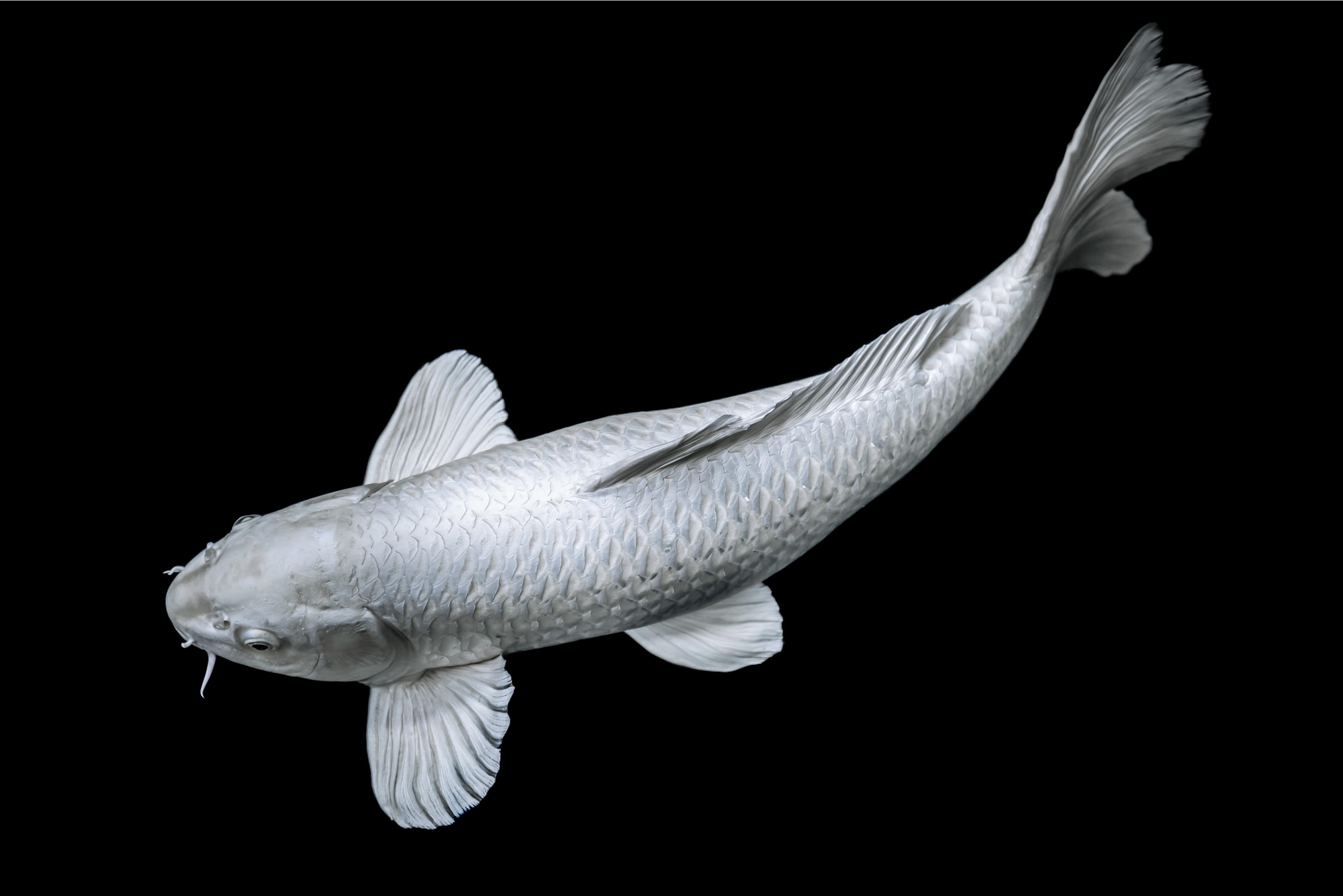}
\includegraphics[width=3.8 cm]{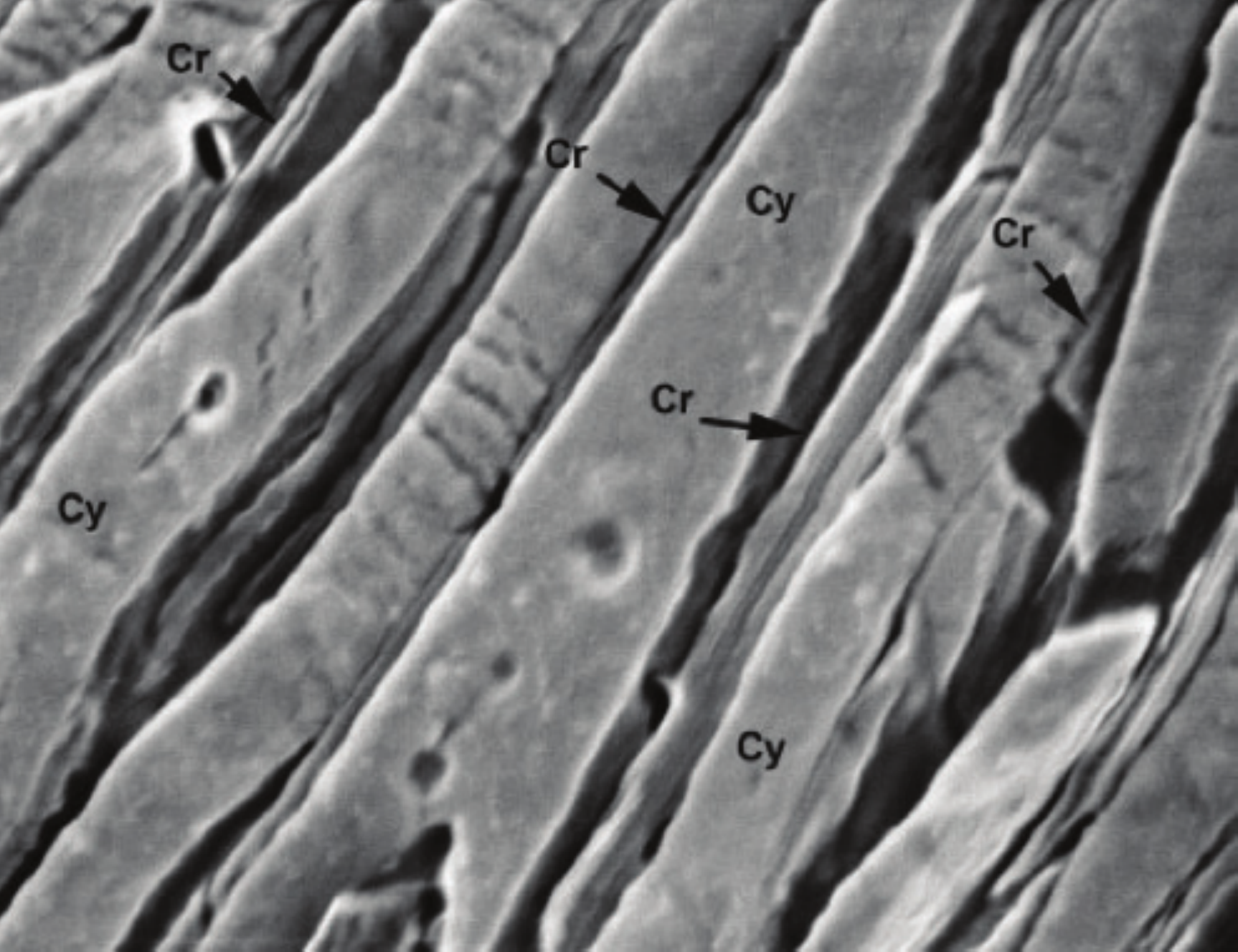}
\end{tabular}
 \caption{\label{Koi}  \textit{Left}: A butterfly Koi fish, exhibiting a broadband reflectance spectrum. \textit{Right}:
 A magnified SEM image of an iridophore cell extracted from the skin of Koi fish \cite{Addadi_2010}, well approximated as an alternating arrangement of guanine crystals (Cr) and cytoplasm (Cy). Image reproduced from Ref. \cite{Addadi_2010}, courtesy of Dr. Dan Oron.}

\end{figure}
\end{center}

The magnitude and distribution of ambient light is perhaps one of the most striking differences between terrestrial and oceanic life and has played a fundamental role in the evolution of the vertical segregation of habitats in our oceans \cite{Herring_Book}. In particular, the key factor determining camouflage in the mesopelagic realm (200-1000 m below sea level), essential for survival in the open ocean environments, is the uniformity of light distribution in all lateral directions. Under such conditions,  a nearly transparent body structure (also found in upper ocean species and flowing rivers) or more practically, a mirror like appearance that mimics transparency, is most advantageous \cite{Denton_1970,Herring_Book}.\\
A mirror like appearance requires a broadband reflectance that reflects all wavelengths within the visible spectrum of light. Sea organisms such as fish achieve this by covering their surface with millions of nearly one dimensional micron sized stacks, comprising of a high refractive index material interspersed with a low refractive index material. Such a one dimensional stack of bilayers is the most basic unit for structural colors in nature \cite{Rayleigh_1917,Kinoshita_Book}. The basic optical structure in fish scales is well approximated by a one dimensional stack of a crystalline organic substance - guanine, with a refractive index approximately $n_g \approx 1.83$, interspersed with layers of watery cytoplasm, with a refractive index approximately $n_c \approx 1.36$, see Fig.\ (\ref{Koi}) (bottom) for a SEM image showing the one dimensional optical stack in Koi fish. The disorder in these structures is manifest in the randomly varying thicknesses of either or both of these materials. As discussed in Refs. \cite{jordan1, jordan2}, Anderson localization provides a natural framework to study the reflection spectra of these disordered stacks.

\begin{center}
\begin{figure}[h]
\begin{tabular}{c}
\includegraphics[width=8 cm]{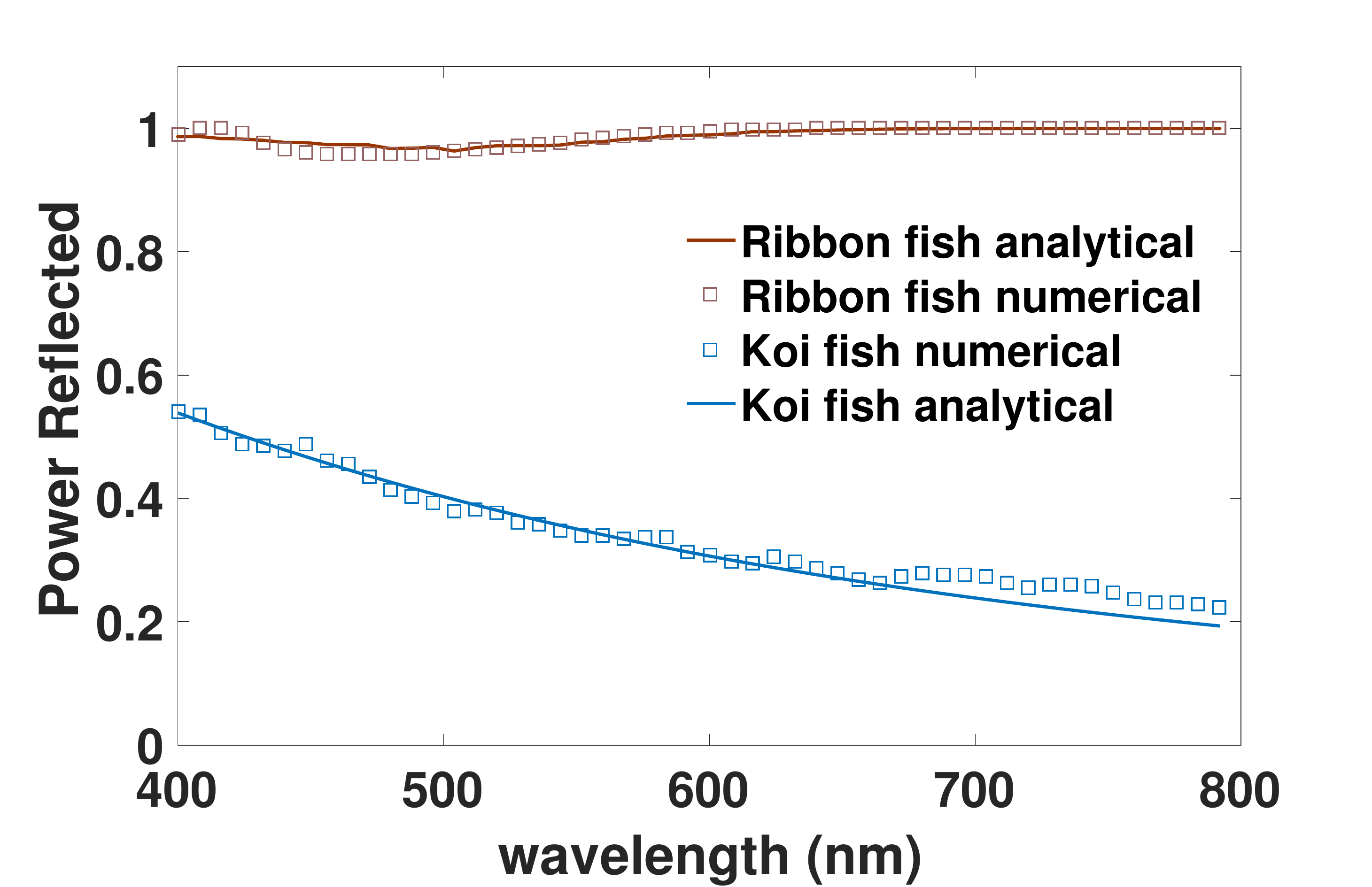} 
\end{tabular}
 \caption{\label{Ribbon_Koi} The power reflected by a one dimensional disordered stack of guanine crystals ($n_g=1.83$) interspersed with layers of cytoplasm ($n_c=1.36$). The blue curves correspond to parameters of the Koi fish with thickness of cytoplasm layers Gaussian distributed between 230 nm and standard deviation 94 nm and thickness of guanine layer nearly constant at 19 nm for a total of 64 layers. For comparison, the red curves correspond to parameters of the Ribbon fish with thickness of cytoplasm layers uniformly distributed between 150$\pm$75 nm and thickness of guanine layer uniformly distributed between 110$\pm 55$ nm for a total of 200 layers. Here, the square symbols are obtained numerically by ensemble averaging the logarithm of the transmitted power ($T$) over 1000 samples to evaluate the reflected power $1-\text{exp}(\langle\text{log}T\rangle)$  while the dashed curves are the analytical expression given in Eq.\ (\ref{Power_R}).  The reflection spectrum corresponds to normal incidence.}

\end{figure}
\end{center}

The guanine-cytoplasm stack in Koi fish consists of cytoplasm layers whose thicknesses are normally distributed with a mean $d_c=230$ nm and standard deviation $\sigma _c=94$ nm, while the guanine layers are also normally distributed with a mean $d_g=19$ nm and variation of $\sigma _g=5$ nm \cite{Addadi_2010}. Consequently, we will make the working assumption that the guanine-cytoplasm stack in Koi fish may be well approximated by a one-dimensional system of scatterers (guanine) placed randomly in a medium of cytoplasm implying that the transmission spectrum is well approximated by Eq. \ (\ref{Power}) with a localization length given in Eq. \ (\ref{Localization_Length_Zeta}).

For comparison, we also discuss the case of Ribbon fish, which is another fish that displays silver coloration. The guanine-cytoplasm stack in Ribbon fish consists of cytoplasm layers whose thicknesses are uniformly distributed with mean $d_c=150$ and standard deviation $\sigma _c=75$ nm, while the thickness of the guanine layer are also uniformly distributed with mean $d_g=110$ and standard deviation $\sigma _g=55$ nm \cite{McKenzie_1995}. Defining the perturbation parameter $\epsilon=\frac{2\pi}{\lambda}\sigma$, we find that for a typical $\lambda= 600 $nm and standard deviation $\sigma\approx 50$ nm, $\epsilon \approx 0.5$. As discussed and derived in reference \cite{Makarov_2009}, we can therefore consider this system weakly disordered and by expanding the transfer matrices $\mathcal{M}_i$ to second order in $\epsilon$, we obtain the localization length (Note, the corresponding perturbation parameter for the case of Koi fish evaluates to $\epsilon\approx 1.1$ and consequently, the weak disorder perturbation expansion is not justified there and refer to this regime as the strongly disordered regime)--
\begin{flalign}
l_r = (d_g+d_c)\left[\frac{\text{sin}^2\left(\gamma\lambda\right)}{4\pi^2\left(n_g^2{\sigma^2 _g}\text{sin}^2(\delta _c)+n_c^2{\sigma^2_c}\text{sin}^2(\delta _g)\right)\alpha^2}\right]. \label{Loc_Ribbon}
\end{flalign}
Here, $n_{g,c}$ are the refractive indices for the guanine(g), cytoplasm(c) layers, $d_{c,g}$ their mean thicknesses, $\delta_{c,g}=\frac{2\pi}{\lambda}d _{c,g}$ the mean optical phases, while ${\sigma _{g,c}}$ are their respective standard deviations. We have defined the bare reflection coefficient as
\begin{flalign}
 r=\left|\frac{n_g-n_c}{n_g+n_c}\right| \label{refl_coeff}
\end{flalign}
in terms of which,  $\alpha = \frac{2r}{1-r^2}$, while ${\gamma}$ is the mean complex Bloch wave-vector, see Appendix A, Eq.\ (\ref{bloch}). 


In Fig.\ (\ref{Ribbon_Koi}), we have compared the numerically evaluated reflection spectrum (dotted curves) for normal incidence against the analytic expression (solid curves) given by
\begin{flalign}
R=1-e^{-\frac{2L}{l_{r,k}}}, \label{Power_R}
\end{flalign}
where, $L$ is the system size and $l_{r,k}$ are the localization lengths for the Ribbon fish (red curves) and for the Koi fish (blue curves) respectively and find a reasonably good agreement over the visible spectrum. 

%

\section{Design of narrow passband disordered acoustic metamaterials}

In the previous section, we discussed how Anderson localization could quantitatively explain the main mechanisms of the origin of silver coloration in two species of fish. These constitute examples of natural occurring disordered systems where disorder has been harnessed to achieve broadband reflectance, important for the survival of many species of fish in their environment. In this section, we will discuss  how the results in Eq.\ (\ref{Power}) can be used to open a narrow band of transmittance in an otherwise reflecting layer thereby utilizing disorder to design narrow passband transmittance filters. Since Anderson localization is a generic wave phenomena, we will discuss the design of narrow pass band transmittance for the acoustic case. 
However, the same idea may be applied for the design of optical or microwave filters.

\begin{figure}[h]
\begin{tabular}{c}
\includegraphics[width=7.5cm]{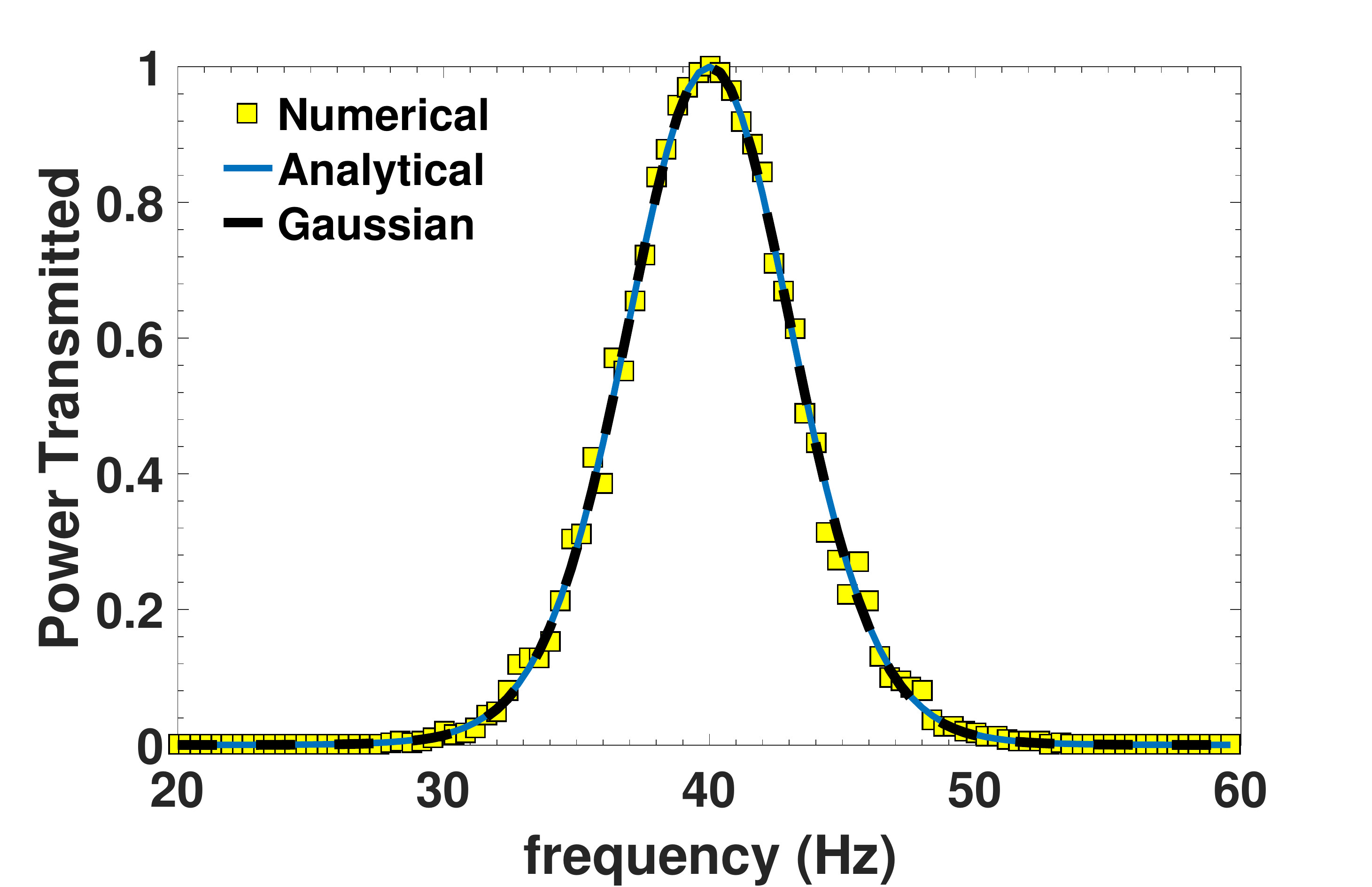} \\ \includegraphics[width=7.5cm]{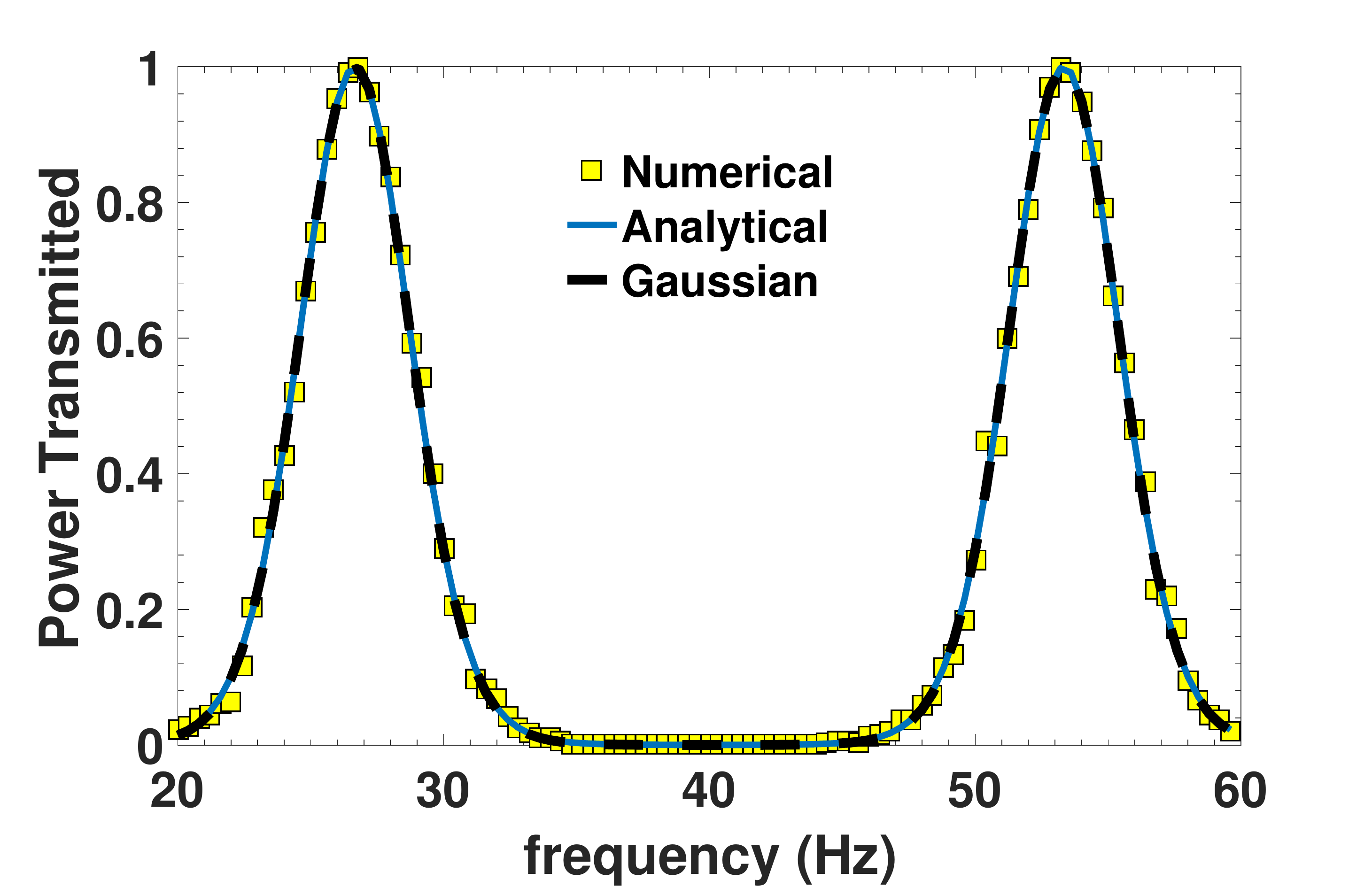} \\
\end{tabular}
\caption{\label{TransMatrix_Disorder} The average power transmitted as a function of frequency through a disordered meta-material consisting of a periodic on average alignment of 100 bilayers where the thickness of one of the layers (scatterer) is held constant at a value $d_2$, while the thickness of the disordered layer is drawn from a uniform distribution with mean value $\langle d_1\rangle=0.5$ m and standard deviation $\sigma=0.4$ m and we have chosen material sound speeds as $c_1=5.33$m/s and $c_2=4$m/s and linear densities as $\rho _1=0.225$Kg/m and $\rho _2=0.4$Kg/m. As we change the thickness of the fixed length scatterer $d_2$, the channel allows the transmission of harmonics $f=\frac{n}{2}\frac{c_2}{d_2}$, where $n$ is a positive integer. For illustration, we have chosen two values: $d_2=0.05$ m (top) and $d_2=0.075$ m (bottom). The symbols (yellow square) correspond to numerical data obtained after ensemble averaging 100 realizations of the logarithm of the power transmitted obtained by calculating the product of transfer matrices comprising the disordered meta-material, the solid curve (blue) corresponds to the analytical expression derived in Eq.\ (\ref{Power}) while the dash-dash curve (black) corresponds to the Gaussian approximation given in Eq.\ (\ref{gaussian}).}
\end{figure}

In Fig.\ (\ref{TransMatrix_Disorder}), we compare the analytical results (dashed line) in Eq.\ (\ref{Power}) with the ensemble averaged power transmitted calculated numerically  (square symbols) by evaluating the product of random transfer matrices comprising the disordered meta-material. Here, the thickness of the disordered layer is drawn from a uniform distribution with mean value $\langle d_1\rangle=0.5$ m and standard deviation of $\sigma=0.4$ m.

As we change the thickness of the fixed length scatterer $d_2$, the frequencies which are transmitted by the disordered meta-material are the ones which make the localization length in Eq.\ (\ref{Localization_Length_Zeta}) infinity and these occur when $\sin^2 \delta _2=0$, i.e., at frequencies $f =\frac{n}{2}\frac{c_2}{d_2}$, where $n$ is a positive integer, $f$ is frequency in Hertz and $c_2$ is the speed of sound through the scatterer. The plots shown in Fig.\ (\ref{TransMatrix_Disorder}) correspond to $d_2=0.05$m (top) and $d_2=0.075$m (bottom). For instance, if $d_2=0.075$m, $c_2=4$m/s, the frequencies which are transmitted within the range 20-60 Hz are $f_1 = n\frac{80}{3}$ for $n=1,2$, i.e., harmonics of $f_0=\frac{80}{3}$ Hz. As can be observed in Fig.\ (\ref{TransMatrix_Disorder}), we have two peaks around $f_1=26.6$ Hz (n=1) and around $f_2=53.3$ Hz (n=2). All other frequencies within the range $20-60$ Hz are reflected. The material parameters chosen here or other values of sound speeds could for instance be realized using metamaterials with tuneable elastic constants, such as a linear chain of elastic spheres \cite{Nesterenko} or quasi-one dimensional spring networks \cite{Ulrich_2013}. 

\begin{figure}
\includegraphics[width=7.5cm]{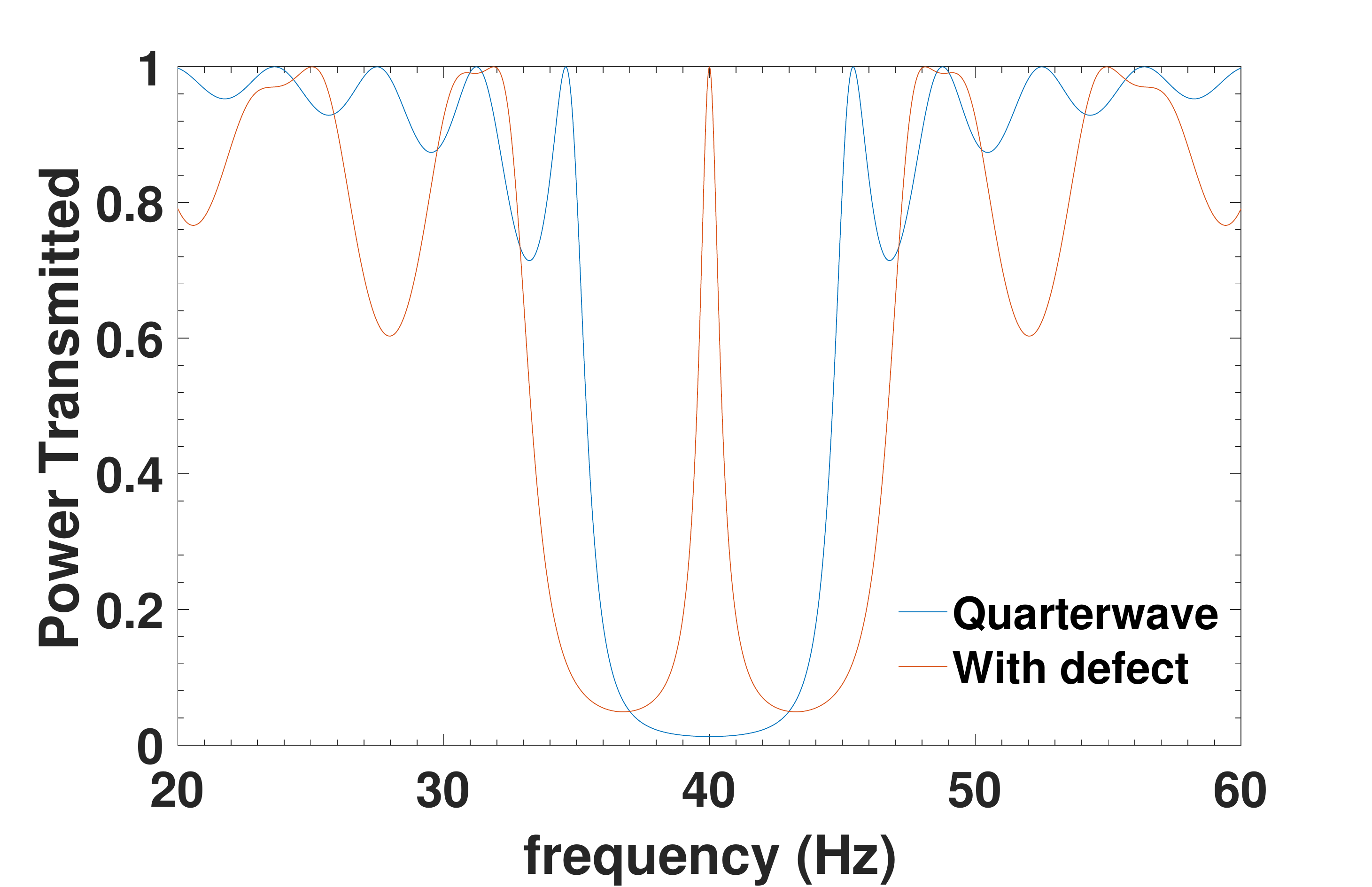}
\caption{\label{TransOrdered} The blue curve shows the power transmitted as a function of frequency through a periodic ordered bilayer (10 bilayers) channel formed from two materials with sound speeds $c_1=5.33$m/s, $c_2=4.0$m/s and with thicknesses tuned to quarter wavelength, while the red curve shows the power transmitted through a similar ordered bilayer (10 bilayers) after the introduction of a half wavelength cavity tuned to give a transmittance peak near the center of the reflection band.}
\end{figure}

Next, we compare the results of the disordered setup with those of an ordered system with a defect-- effectively creating a cavity. Fig.\ (\ref{TransOrdered}) shows the transmission spectrum of a purely ordered structure (blue curve) with layer thicknesses tuned to quarter wavelength creating a reflective layer centered around $f=40$ Hz. This scenario can be solved exactly, and for completeness we derive the transmission of this setup in Appendix A. Upon insertion of a half-wave cavity in the otherwise quarter wavelength reflecting stack \cite{Orfanidis_Book, asano}, a narrow Lorentzian shaped transmission peak is created (see Fig. (\ref{TransOrdered}) red curve) near the middle of the reflection band. One approach to analyze the transmission of this setup is to ``lump" the all multilayers leading up to the defect into one element with reflection $\tilde{r}$ and transmittance $\tilde{t}$, which will also be the reflection and transmittance of the layers following the defect. Now the transmission coefficient is mathematically equivalent to that of a Fabry-Perot setup, and can readily be evaluated to give \cite{AJP}:
%

\begin{flalign}
T \approx \frac{1}{1 + \frac{4R}{(1-R)^2}\sin^2(\delta_L)},
\end{flalign}
with $R \equiv |\tilde{r}|^2$ and $\delta_L$ the phase accumulated in the defect layer. For wavelengths within the stopbands of the ordered multilayer, $1-R$ will be exponentially small in the number of layers, and we find:

\begin{flalign}
T \approx \frac{1}{1 + 4 e^{N \Gamma }\sin^2(\delta_L)},
\end{flalign}
with $\Gamma = i \gamma $ is real and positive in the stopband region, see Eq. (A13). The structure will thus show transmittance peaks when $n_Ld=n\frac{\lambda}{2}$, i.e., at integer multiples of half-wavelength. In contrast to Eq. (\ref{gaussian}), this has a Lorentzian form, as is the typical case for resonances.


What, if any, are the design advantages of the disordered analog of the half-wave cavity? Even though the disordered design is not inherently superior to the ordered half-cavity effect in terms of performance, it has the advantage that it might be easier to fabricate, since if we have a way to synthesize layers of constant thickness, the Onsager mechanism of creating order based on entropic considerations alone, could help self-organize a layered structure \cite{Onsanger_1949}. Such ordering may also arise due to flow, as is the case in rheoscopic fluids \cite{rheoscopic}.

To conclude, we have studied Anderson localization of waves in a one dimensional disordered meta-material comprising of fixed length scatterers placed randomly along a homogenous medium. As an interplay between order and disorder, we have identified a new regime of strong disorder where the localization length becomes independent of the amount of disorder but depends on the frequency of the wave excitation and on the properties of the fixed length scatterer. As an example of a naturally occurring disordered system, we have compared our analytical results with numerical results evaluated using experimental obtained parameters for two species of fish, and used the formalism of Anderson localization to explain the emergence of their silver structural coloration. Further, we have discussed the design of a disordered narrow pass band acoustic filter which gives performance analogous to a half-wave cavity inserted in a quarter wavelength reflecting stack. We believe our results could stimulate further analysis and experimental work harnessing disorder to engineer useful materials.

\emph{Acknowledgments.} We thank Dan Oron, Mathias Kolle, Caleb Cook, Kevin Zhou, Vinny Manoharan, William Rogers and Samiksha Joshi for useful discussions. AA thanks the Kavli Foundation and the Harvard MRSEC program of the NSF (DMR 14-20570) for their support.

\clearpage

\appendix

\section{Review of transfer matrix formulation for an ordered meta-material}

Consider a dielectric bilayer composed of materials with refractive indices $n_1,n_2$ and thicknesses $d_1,d_2$ respectively, stacked along the $z$-direction. The total electric field in any layer is a superposition of forward and reflected electric fields,
\begin{align}
\tilde{E}_1= E^{+}_{1}e^{i\bold{k}_1\cdot\bold{r}} + E^{-}_{1}e^{i\bold{k}'_1\cdot\bold{r}}.
\end{align}
Here, $E^{+,-}_{1}$ are the electric field amplitudes where the superscript $+,-$ denote forward and backward traveling waves respectively, and $\bold{k}_1,\bold{k}'_1$ are the wave vectors specifying the direction of propagation of incident and reflected waves. Likewise, the total electric field in layer-2 is,
\begin{align}
\tilde{E}_2= E^{+}_{2}e^{i\bold{k}_2\cdot\bold{r}} + E^{-}_{2}e^{i\bold{k}'_2\cdot\bold{r}}.
\end{align}

Next, assume that the incident wave makes an angle $\theta _1$ with the normal to the interface situated at $z=d_1$. Then, from the boundary conditions we obtain $|k_1|=|k'_1|$, $|k_1|\text{sin}\theta _1=|k_2|\text{sin}\theta _2$, where, $\theta _2$ is the angle of refraction. For a plane polarized wave we obtain the following constraints between the electric field amplitudes \cite{Griffiths_Book}
\begin{align}
\frac{1}{\alpha _1}E^{+}_{1}e^{i\delta _1} + \frac{1}{\alpha _1}E^{-}_{1}e^{-i\delta _1} = E^{+}_{2} + E^{-}_{2},\\
\frac{1}{\beta _1}E^{+}_{1}e^{i\delta _1} - \frac{1}{\beta _1}E^{-}_{1}e^{-i\delta _1} = E^{+}_{2} - E^{-}_{2},
\end{align}
where, $\bold{r}=d_1\hat{z}$ (along the direction of the stack) and we have defined $\delta _1=|k_1|d_1\text{cos}\theta _1$. Here, $\alpha _1=\frac{\text{cos}\theta _2}{\text{cos}\theta _1}$ and $\beta _1=\frac{\mu _1n_2}{\mu _2n_1}$. It is convenient to express these relations in the form of a transfer matrix relating the electric field amplitudes to the left and right of a dielectric layer-
 \begin{align*}
\begin{pmatrix}
E^{+}_{2}\\
E^{-}_{2}\\
\end{pmatrix}=\begin{pmatrix}
\frac{\alpha _1+\beta _1}{2\alpha _1\beta _1}e^{i\delta _1} & \frac{\beta _1-\alpha _1}{2\alpha _1\beta _1}e^{-i\delta _1} \\
\frac{\beta _1-\alpha _1}{2\alpha _1\beta _1}e^{i\delta _1} & \frac{\alpha _1+\beta _1}{2\alpha _1\beta _1}e^{-i\delta _1} \\
\end{pmatrix}
\begin{pmatrix}
E^{+}_{1}\\
E^{-}_{1}\\
\end{pmatrix}
\end{align*}
Note, due to energy conservation (we have assumed no absorption) and time reversal invariance, the transfer matrix has a real trace and unit determinant \cite{Wolf_Book}. Consequently, any product of these matrices will also have unit determinant.

Likewise, for the wave incident from medium-2 to medium-1, $\beta _2=\frac{1}{\beta _1}$, $\alpha _2=\frac{1}{\alpha _1}$ and $\delta _2=|k_2|d_2\text{cos}\theta _2$ and we find the matrix relation,
 \begin{align*}
\begin{pmatrix}
E^{+}_{3}\\
E^{-}_{3}\\
\end{pmatrix}=\begin{pmatrix}
\frac{\alpha _1+\beta _1}{2}e^{i\delta _2} & -\frac{\beta _1-\alpha _1}{2}e^{-i\delta _2} \\
-\frac{\beta _1-\alpha _1}{2}e^{i\delta _2} & \frac{\alpha _1+\beta _1}{2}e^{-i\delta _2} \\
\end{pmatrix}
\begin{pmatrix}
E^{+}_{2}\\
E^{-}_{2}\\
\end{pmatrix}
\end{align*}

The overall transfer matrix for the bi-layer is now,
 \begin{align}
\mathcal{M}=\begin{pmatrix}
A & B \\
B^* & A^{*} \\
\end{pmatrix},
\end{align}
where,
\begin{align}
A &=& \frac{1}{1-r^2}\left(e^{i(\delta _1+\delta _2)}-r^2e^{i(\delta _1-\delta _2)}\right),\\
B &=& \frac{2ir}{1-r^2}e^{-i\delta _1}\text{sin}\delta _2,
\end{align}
where, $r=\frac{\beta _1-\alpha _1}{\beta _1+\alpha _1}$. (An identical calculation for a wave that is polarized perpendicular to the plane leads to the same form for the transfer matrix, with the replacement $r=\frac{\alpha _1\beta _1-1}{\alpha _1\beta _1+1}$).

Next, we define the product of $N-1$ matrices as,$\bold{M_N}=\mathcal{M}_{N-1}\mathcal{M}_{N-2}\cdot\cdot\mathcal{M}_{1}$ to obtain the overall relation between the incident wave amplitude and wave amplitude after traversing $N$ bi-layers-
 \begin{align}
\begin{pmatrix}
E^{+}_{N}\\
E^{-}_{N}\\
\end{pmatrix}=\bold{M}_N
\begin{pmatrix}
E^{+}_{1}\\
E^{-}_{1}\\
\end{pmatrix},\label{ANproduct}
\end{align}

Upon normalizing the columns in Eq.\ (\ref{ANproduct}) with respect to the incident wave amplitude, $E^{+}_{1}$, we find,
 \begin{align}
\begin{pmatrix}
\frac{E^{+}_{N}}{E^{+}_{1}}\\
\frac{E^{-}_{N}}{E^{+}_{1}}
\end{pmatrix}=\bold{M_N}
\begin{pmatrix}
1\\
\frac{E^{-}_{1}}{E^{+}_{1}}
\end{pmatrix}.\label{Nproduct2}
\end{align}
By definition, the (complex) transmitted amplitude is $\tau=\frac{E^{+}_{N}}{E^{+}_{1}}$ and the complex reflection amplitude is $\rho=\frac{E^{-}_{1}}{E^{+}_{1}}$, while there is no backward propagating wave at the end of the $N$ layers, i.e., $\frac{E^{-}_{N}}{E^{+}_{1}}=0$. Solving the matrix equations, we find
\begin{eqnarray*}
\tau = \bold{M}_N(1,1)+ \bold{M}_N(1,2)\rho,\\
0 = \bold{M}_N(2,1)+ \bold{M}_N(2,2)\rho,
\end{eqnarray*}
where, in addition, $\bold{M}^*_N(1,1)=\bold{M}^*_N(2,2)$ and $\bold{M}^*_N(1,2)=\bold{M}^*_N(2,1)$. Solving these, we obtain,
\begin{align}
\rho = -\frac{\bold{M}_N(2,1)}{\bold{M}_N(2,2)} \label{reflectance}
\end{align}
and
\begin{align*}
\tau =  \frac{\bold{M}_N(2,2)\bold{M}_N(1,1)-\bold{M}_N(2,1)\bold{M}_N(1,2)}{\bold{M}_N(2,2)}.
\end{align*}
Since det$\{\bold{M}_N\}=1$, the above expression simplifies to
\begin{align}
\tau =  \frac{1}{\bold{M}_N(2,2)}.  \label{transmittance}
\end{align}
Therefore, the overall power transmitted is $|\tau|^2=T = \frac{1}{|\bold{M_N}[1,1]|^2}=\frac{1}{|\bold{M_N}[2,2]|^2}$. By energy conservation, the power reflected is $R=1-T$. \\

For the ordered case, where all the layers have the same thicknesses, we have the following analytic expression for the overall power transmitted \cite{Makarov_PRB}
\begin{align}
T_N = \frac{1}{1 + |\mathcal{M}(1,2)|^2\left(\frac{\text{sin}(N\gamma)}{\text{sin}(\gamma)}\right)^2}, \label{Power_Ordered}
\end{align}
where, $\gamma$ is the complex Bloch wave vector that satisfies the condition,
\begin{align}
2\text{cos}\gamma= \text{Trace}\{\mathcal{M}\}.\label{bloch}
\end{align}

An analogous derivation follows for one dimensional mechanical displacement fields (acoustic) where we identify the parameter $\beta = \frac{\rho _2c_2}{\rho _1c_1}$ where, $\rho _{1,2}$ are the densities of the two layers comprising the bilayer and $c _{1,2}$ the corresponding speed of sound in the respective layers.

\section{Alternate Derivation of localization length}\label{App: AppendixB}

In  this section, we re-derive the localization length Eq.\ (\ref{Localization_Length_Zeta}) by following the derivations in reference \cite{Berry_1996}, albeit, keeping track of an additional wavelength dependent factor which in this case, comes from the barrier layer having a nearly constant thickness. Since the random variations occur in only the background medium, we write the transfer matrix given in Eq.\ (\ref{transMatrix}) for the $n-th$ bi-layer as,
 \begin{align*}
\mathcal{M}_n &=& \frac{|E|e^{-i\delta _n}}{1-r^2}\begin{pmatrix}
\frac{E}{|E|}z_n & \frac{\Sigma}{|E|} \\
\frac{\Sigma^*}{|E|}z_n & \frac{E^*}{|E|} \\
\end{pmatrix},\\
 &=& \frac{|E|e^{-i\delta _n}}{1-r^2}\tilde{\mathcal{M}}_n.
\end{align*}
where, $E= e^{i\delta _2}-r^2e^{-i\delta _2}, \Sigma = 2ir\text{sin}\delta _2$ and $z_n=e^{2i\delta _n}$. Here, $\delta _2$ is a constant phase across the barrier layer (no random variation), while $\delta _n$ is the random phase accumulated as the wave propagates through the background layer after $n-$ bilayers.

For a stack of $N-$ bi-layers, we use the following definition for the Lyuponov exponent (which is proportional to the inverse Anderson localization length)
\begin{align}
\zeta _b = 2\lim _{N\rightarrow\infty}\frac{\text{Re}\langle\text{log}\text{Trace}\{\bold{M}_N\} \rangle}{N}, \label{Lyuponov}
\end{align}
whereby, we need to evaluate the expression,
\begin{align}
 \text{Re}\left\langle\text{log}\left(\left(\frac{|E|}{1-r^2}\right)^Ne^{-i(\delta _1+\delta _2\cdot\cdot+\delta _n)}\text{Trace}\{\mathcal{\tilde{M}}_n\cdot\cdot\mathcal{\tilde{M}}_1\}\right)\right\rangle.  \label{Lyp_R}
\end{align}
Upon taking the real part of Eq.\ (\ref{Lyp_R}), the exponential factor containing random phases drops out. The expectation value can now be evaluated by assuming that due to strong disorder, the phases $\delta _n$ take all possible values from $0-2\pi$. Thus, we convert the phase integral into a contour integral in terms of the complex random variable $z_n$(where, $|z_n|=1$ over the unit circle), to obtain the expression \cite{Berry_1996},
\begin{align}
\zeta _b &=& 2\text{log}\left(\frac{|E|}{1-r^2}\right) + 2\lim _{N\rightarrow\infty}\frac{1}{(2\pi i)^N}\oint \frac{dz_1}{z_1}\cdot\cdot\oint\frac{dz_n}{z_n}\\
               & \times& \text{log}\text{Trace}\left[\begin{pmatrix}
\frac{E}{|E|}z_1 & \frac{\Sigma}{|E|} \\
\frac{\Sigma^*}{|E|}z_1 & \frac{E^*}{|E|} \\
\end{pmatrix}
\cdot\cdot
\begin{pmatrix}
\frac{E}{|E|}z_n & \frac{\Sigma}{|E|} \\
\frac{\Sigma^*}{|E|}z_n & \frac{E^*}{|E|} \\
\end{pmatrix}
\right].
\end{align}
The contour integrals on the right-hand side are given by the residues at $z_k=0$ ($k=1\cdot\cdot n$) and therefore vanish \cite{Berry_1996}. Consequently, the exponent is
\begin{align}
\zeta _b = \text{log}\left(1+\frac{4r^2}{(1-r^2)^2}\text{sin}^2(\delta _2) \right),\label{Lyp_Koi}
\end{align}
from where the localization length in the strongly disordered regime is
\begin{align}
l _b = \frac{d}{\zeta _b},\label{Loc_Koi}
\end{align}
recovering Eq.\ (\ref{Localization_Length_Zeta}) given in the main text.


\begin{thebibliography}{10}
\bibitem{Log_note_2} Regarding the notation here, we are not implying that mean($T$) is the same as $\exp(\text{mean}(\text{log}T)))$, rather, Eq.\ (\ref{Power}) is simply to suggest that the value of transmission we expect to observe is $\exp(\text{mean}(\text{log}T))$.  
\end{thebibliography}
\end{document}